\documentclass[apj]{emulateapj}      

\renewcommand{\baselinestretch}{1.4}

\shortauthors{Kasun \& Evrard} 
\shorttitle{Galaxy Cluster Shapes}

\def \mpc {{\rm\ Mpc}}

\def \etal {et al.\ }

\def \Ho {{\rm\ H_{o}}}
\def \kmsmpc {{\rm km\ s^{-1}\ Mpc^{-1}}}

\def \msol {{\rm M}_\odot}

\def \hinv {\hbox{$\, h^{-1}$}}

\def \se {\!=\!}
\def \sims {\sim \!}
\def \ssim {\! \sim \!}
\def \ssimeq {\! \simeq \!}

\def \spropto {\! \propto \!}
\newcount\itemno
\def \ino { \the\itemno\global\advance\itemno by 1}
\def \rtwoh {\hbox{$r_{200}$}}
\def \Mtwoh {\hbox{$M_{200}$}}

\def \lcdm {\hbox{$\Lambda$CDM}}
\def \tcdm {\hbox{$\tau$CDM}}
\def \ctilde {\tilde{c}}
\def \btilde {\tilde{b}}

\def \caeq {{$\tilde{c}$}}
\def \baeq {{$\tilde{b}$}}
\def \Mcut {3 \times 10^{14} \hinv \msol}

\begin{document}

\title{SHAPES AND ALIGNMENTS OF GALAXY CLUSTER HALOS \\ {\hskip 1.0cm} \\}
\shorttitle{Galaxy Cluster Shapes}
\author{S.F. Kasun \and A.E. Evrard}

\affil{Departments of Physics and Astronomy  and
Michigan Center for Theoretical Physics \\
University of Michigan, Ann Arbor, MI 48109-1120 USA } 

\email{skasun@umich.edu, evrard@umich.edu}

\slugcomment{DRAFT, submitted to ApJ \today}

\begin{abstract}

We present distribution functions and spatial correlations of the
shapes of dark matter halos derived from Hubble Volume simulations of a  
\lcdm\  universe.  We measure both position
and velocity shapes within spheres encompassing mean density 200 times
the critical value, and calibrate small-N systematic errors
using Poisson realizations of isothermal spheres and higher resolution
simulations.  For halos more massive than $3 \times 10^{14} \hinv
\msol$, the shape distribution function peaks at (minor/major,
intermediate/major) axial ratios of $(0.64,0.76)$ in position, and is 
rounder in velocity, peaking at $(0.72,0.82)$.  Halo shapes are
rounder at lower mass and/or redshift; the mean minor axis ratio in
position follows $\langle c/a \rangle (M,z) = c_{15,0} [1-\alpha
\ln(M/10^{15}\hinv\msol)] (1+z)^{-\varepsilon}$, with 
$c_{15,0} \se 0.631 \pm 0.001$, $\alpha \se 0.023 \pm 0.002$ and
$\varepsilon \se 0.086 \pm 0.004$.
Position and velocity principal axes are well aligned
in direction, with median alignment angle $22^\circ$, and the
axial ratios in these spaces are correlated in magnitude.  
We investigate mark correlations of halo pair orientations using two
measures:  a simple scalar product shows $\ge 1\%$ alignment extending
to $30 \hinv \mpc$ while a filamentary statistic exhibits non-random 
alignment extending to scales $\sims 200 \hinv \mpc$, ten times the
sample two-point correlation length and well into the regime of
negative two-point correlation.  Cluster shapes are unaffected by the
large-scale environment; the shape distribution of supercluster
members is consistent with that of the general population.
\end{abstract}

\smallskip

\medskip
\setcounter{footnote}{0}

\section{INTRODUCTION}
Clusters of galaxies signal the largest gravitationally bound dark matter
halos in the universe.  They are mildly aspherical systems that tend
to be aligned by mergers directed by interconnecting filaments in the cosmic
web.  

Since the early days of extragalactic astronomy, it has been apparent
that clusters are generally elongated on the sky.  Flattening of these
clusters due to rotation was ruled out \cite{ill77} and many assumed
that the flattening was due to gravitational instabilities expected 
in the top-down scenario of Zel'dovich (1978) and 
Doroshkevich \etal (1978).
The groundbreaking work of Carter and Metcalfe
(1980) showed that the aspherical shape of a cluster was connected with the
velocity anisotropy of the orbits of cluster galaxies.  Binney and
Silk (1979) proposed that this anisotropy was due to 
tidal distortion from neighboring large-scale structure. 

Binggeli (1982) was the
first to investigate alignments of close cluster pairs.  He studied
44 Abell clusters and found that galaxies separated by less than 30
Mpc show a strong alignment with each other and that the
orientation of a cluster was dependent on the distribution of
surrounding clusters.  
Other observational claims of cluster alignments have been made,
both with nearest neighbors and with other 
clusters in the same supercluster \cite{Iwest89, IIwest89, rhee92, rich92, pli94}.   

With some exceptions (for example, Struble and Peebles (1985) or
Rhee and Katgert (1987)), most work in the literature \cite{det95, west95, Iwest89, rhee92, onu00} confirms Binggeli's 
original results.  West (1989a) uses 48 Abell superclusters and
finds a tendency for clusters within $60 \hinv \mpc$ to be
aligned.  Plionis (1994) measures alignments for 637 Lick clusters 
and finds strong alignments up to $15 \hinv \mpc$ with weaker alignments
out to $60 \hinv \mpc$.  West et al. (1995) finds a marked anisotropy
for Einstein clusters extending out to $10 \hinv \mpc$.  

Simulations have shown that spatial alignments, intrinsically predicted from 
the 'top-down' model \cite{zel70}, are also seen in bottom-up
(cold dark matter) scenarios, where halos form by 
mergers of smaller structures organized along filaments
\cite{str94, west91, west95}. Simulations also show that the
orientation  
of the major axis of a galaxy cluster is aligned with the direction of the 
last major merger event \cite{van93, spl97}.  Since
cluster alignments 
appear to be a generic outcome of gravitational 
instability, their use as a discriminant of cosmological models
requires careful calibration \cite{onu00}.  

In the halo model description of non-linear structure (Cooray and
Sheth 2002), all matter is contained in bounded, spatially correlated
regions (the halo population) that span a spectrum of
sizes.  Most instances of this model assume spherically symmetric
halos, but more precise versions will need to take into account the
spectrum of halo shapes, including detailed internal structure (Jing \&
Suto 2002), as well the spatial correlation of the shapes of
neighboring halos (Jing 2002; Faltenbacher \etal 2003).  

In this paper, we report measurements of shape statistics,
including spatial (or {\sl mark}) 
correlations of alignments, derived from large samples of 
massive dark matter halos extracted from Hubble Volume simulations.  We
investigate two flat-metric cosmologies (\lcdm\ and \tcdm) dominated
by vacuum energy and dark matter, respectively.  We focus on the
former, more empirically satisfying model, but show results for the
latter for comparative purposes.  Section~2 provides details of the 
simulations and describes our method of finding principal axis
orientations and magnitudes for mass-limited halo samples.  In \S~3, we
use Poisson realizations of isothermal spheres to estimate 
systematic error in mean shape measurement due to shot noise.  We then 
present axial ratio distribution functions for mass-limited samples,
and characterize the dependence of halo shape on mass and redshift.  
Section~4 presents mark correlations of \lcdm\ cluster alignments and briefly
examines the role of supercluster membership on shape. A
final section reviews our conclusions.  

Unless explicitly stated otherwise, the halo mass $M$ used throughout
this paper is a critically-thresholded, spherical overdensity mass
($\Mtwoh$) expressed in units of $10^{15} \hinv \msol$, with $h \se
\Ho/100 \kmsmpc$.  

\section{CLUSTER SHAPES FROM SIMULATIONS}

We use two sets of simulations to investigate dark matter halo shapes.  The
large Hubble Volume (HV) simulations provide statistical power 
but poor mass resolution, while higher resolution, but smaller volume,
Virgo simulations are used for resolution tests.  

\subsection{Simulations}
The HV simulations are a pair of gigaparticle N-body 
simulations created using a parallel version of the  Hydra N-body
code \cite{mac98}.  Random realizations of two cosmologies
with flat spatial metric are produced: a \lcdm\  model with 
$\Omega_{m}=0.3$, $\Omega_{\Lambda}=0.7$ and power spectrum
normalization $\sigma_{8}=0.9$ and a \tcdm\ model with 
$\Omega_{m}=1$, $\Omega_{\Lambda}=0$, and $\sigma_{8}=0.6$.
The dark matter structure is resolved by particles of mass $2.2 \times
10^{12} \hinv\msol$ within periodic cubic volumes of length 3000 and
$2000 \hinv \mpc$, respectively (see Table \ref{tab:models}).  
We analyze $z\se 0$ and combined sky survey samples of galaxy cluster
halos published in Evrard \etal\ (2002).  The reader is referred to
 that paper for details of the simulations, including the process of
sky survey creation.    


\begin{deluxetable}{lcccccc}
\tablewidth{0pt} \tablecaption{Model Parameters \label{tab:models}}
\tablehead{ \colhead{Model} & \colhead{$\Omega_m$} &
\colhead{$\Omega_\Lambda$} & \colhead{$\sigma_8$} &
\colhead{$z_{init}$} & \colhead{$L$\tablenotemark{a}} &
\colhead{$m$\tablenotemark{b}} } 
\startdata 
\lcdm & 0.3 & 0.7 & 0.9 & 35 & 3000 & $2.25$\\ 
\tcdm & 1.0 & 0.0 & 0.6 & 29 & 2000 & $2.22$\\
\enddata
\vspace* {-0.3truecm} 
\tablenotetext{a}{Cube side length in $\hinv \mpc$.}  
\tablenotetext{b}{Particle mass in $10^{12}\hinv\msol$.}
\end{deluxetable}

For resolution tests, we employ the $256^3$--particle
Virgo simulations of Jenkins \etal (1998).  These simulations have
order-of-magnitude improved mass resolution but much smaller samples
than the HV simulations. 

\subsection{Halo Finding Algorithm}

We define dark matter halos with a 
spherical overdensity (SO) group finder that identifies as 
a halo the set of particles lying within a sphere of size \rtwoh, 
centered on a particle that represents a local density maximum
filtered on a scale of $2 \times 10^{13} \hinv \msol$.  The size measure 
$\rtwoh$ is the radius of the sphere within which the mean density is
200$\rho_{c}(z)$, with $\rho_{c}(z)$ the critical density at
redshift $z$.  The total mass $\Mtwoh$ lying within r$_{200}$ is the
basic order parameter of the halo sample.   
As a result of ongoing merging activity, roughly $7.5\%$ of halos 
are found to have overlapping $\rtwoh$ spheres.  We retain both members
of an overlapping pair in the halo sample.

\subsection{Sky Surveys}

In addition to the traditional mode of fixed proper-time output,
sky survey samples, consisting of data collected along 
the past light cone of hypothetical observers located within the
computational volume, were also generated by the HV simulations.  
We use two octant surveys (PO and NO) that cover $\pi/2$ sterad and 
extend to $z_{max} \se 1.46$ for \lcdm\  and 1.25 for \tcdm, as well as
two full-sky surveys (MS and VS) that reach $z_{max} \se 0.57$ for
\lcdm\  and 0.42 for \tcdm.  The octants sample structure over the last
$74\%$ ($\Lambda$CDM) and $71\%$ ($\tau$CDM) of the age of the
universe, approximately a 10 Gyr look-back time.  Halos in these
surveys are defined using the SO method described above.  

\subsection{Cluster Shapes}

Given the modest resolution of the HV data, we take a simple approach
and estimate the shape of a halo using the moments of the material
within \rtwoh.  

With respect to the center of the halo (defined by the local
gravitational potential minimum), we compute the $3\times3$ symmetric 
tensor $M_{jk}$ 
\begin{equation}
M_{jk} \ =\ \frac{1}{N_h}\ \sum_{\alpha} x_{\alpha j}x_{\alpha k}
\label{eq:inertia} 
\end{equation}  
where $x_{\alpha j}$ is the j$^{th}$ component of the displacement
vector of particle $\alpha$ relative to the halo center and $N_h$
is the number of (equal mass) particles in the halo.  We
diagonalize to find eigenvalues $\lambda_{j}$ and unit
eigenvectors. The sorted eigenvalues  
($\lambda_1 > \lambda_2 > \lambda_3$) and vectors 
define principle axes $a,b,c$, with semi-major axis
$a = \sqrt{\lambda_1}$, intermediate $b \se \sqrt{\lambda_2}$, and
$c \se \sqrt{\lambda_3}$ of a triaxial ellipsoid that approximates the
halo.  We refer to the corresponding unit vector directions as
$\hat{a}$, $\hat{b}$ and $\hat{c}$.  

Velocity moments are
solved for in a similar way, using a mean defined as the center of
mass velocity of the material within $\rtwoh$.  For compactness of
notation, we define intermediate and minor axial ratios as follows
\begin{equation} 
\tilde{b}  \ \equiv \ b/a , 
\label{eq:bovera} 
\end{equation}
\begin{equation} 
\tilde{c} \ \equiv \ c/a. 
\label{eq:covera} 
\end{equation}

We have confirmed that the halo population is 
oriented randomly with respect to the Cartesian coordinates of
the simulation, as required by the cosmological principal.

\section{HALO SHAPE STATISTICS}

We begin this section by estimating the error on shape determinations
due to shot 
noise.  From Monte Carlo realizations of distorted isothermal spheres,
we find that population mean values of $\ctilde$ and $\btilde$ are
accurate to a few percent if $N_h \gtrsim 100$.  We then present the joint 
distribution function $p(\ctilde, \btilde)$ at $z\se 0$, and
investigate joint position and velocity shape statistics.  
We follow with the mass and redshift dependence of the minor axis ratio.   

\subsection{Resolution Tests}

\begin{figure} 
  \centering
  \vspace{-1.6truecm}  
   \epsscale{1.3}
    \plotone{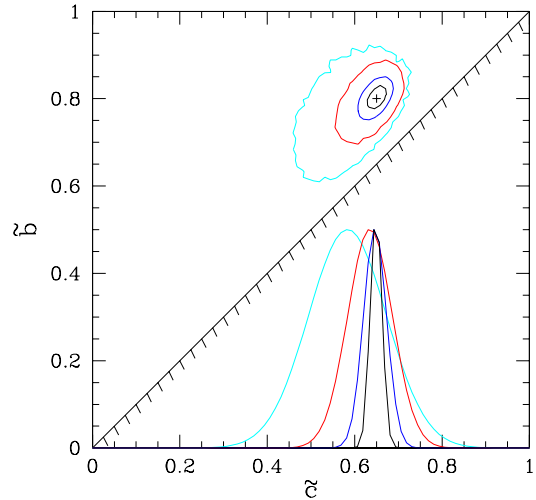}
   \vspace{-1.2truecm} 
   \caption{Estimates of the effect of Poisson noise on measurement
     of minor ($\ctilde$) and intermediate ($\btilde$) axis ratios.
     Halos must lie in the upper left portion of the plot, with 
     spherical objects at $(1,1)$, oblate halos 
     tending to $\btilde \se 1$ and prolate halos lying near 
     the diagonal $\btilde \se \ctilde$.  Contours show $68\%$
     confidence regions of axial ratios measured from 10,000
     random realizations of isothermal
     spheres distorted to mean shape $(0.65,0.80)$, shown by the cross,
     and resolved by 
     (from outer to inner contours) 40, 160, 640, and 2560 particles.
     Histograms in the lower
     right give frequency distributions of the minor axis ratio
     $\ctilde$.  As resolution degrades, the mean is biased to lower
     values and the dispersion grows.  
}
   \label{fig:axial}                                                         
\end{figure}

To calibrate the error in mean shape due to the small numbers
of particles used to resolve clusters in the HV, we carry out the
following resolution test.  Halo models are created using a set of $N_h$ 
particles with an initially spherical, isothermal profile.  To
simulate the mean ellipticity of the HV cluster samples discussed
below, each cluster is then compressed along the $(x,y)$-axes by fixed
amounts (0.65, 0.8).  The particle moments are
calculated and used to estimate the shape of
each cluster in the same manner as for the HV clusters (see
section 2.4).  Generating ensembles at several values of $N_h$, we
calibrate the bias in mean cluster shape as a function of mass
resolution.   

The results for an ensemble of 10,000 clusters at each $N_h$
are presented in Figure \ref{fig:axial}. Mean axial ratio values
$\langle \ctilde 
\rangle$ and $\langle \btilde \rangle$  are presented in Table
\ref{tab:resolution} for $N_h$ ranging from 40 to 2560 particles.  In
Figure \ref{fig:axial}, the measured shape distribution function has a
dispersion ranging from 0.09 at $N_h$=40 to 0.01 at $N_h$=2560, and the
mean is offset from the input values by a bias that scales inversely
with particle number $0.024 (N_h/100)^{-1}$.   We will use 
this calibration to bias-correct the mass-dependent shapes presented  
in \S~3.4. 


\begin{deluxetable}{lcccc}
\tablewidth{0pt} \tablecaption{Resolution Test Values \label{tab:resolution}}
\tablehead{ \colhead{$N_h$} & \colhead{$\langle \tilde{c} \rangle$} &
\colhead{$\sigma_{\tilde{c}}$} & \colhead{$\langle \tilde{b} \rangle$} &
\colhead{$\sigma_{\tilde{b}}$} } 
\startdata 
40   & 0.584 & 0.089 & 0.777 & 0.095\\
160  & 0.635 & 0.051 & 0.794 & 0.061\\
640  & 0.647 & 0.026 & 0.799 & 0.032\\ 
2560 & 0.649 & 0.013 & 0.799 & 0.016\\
\enddata
\vspace* {-0.3truecm} 
\end{deluxetable}

\subsection{Shape Distributions}

To ensure a mean shape measurement biased by less than two
percent, we impose a mass limit $\Mtwoh \ge \Mcut$, corresponding to
133 particles in the HV simulations.  We note that Jing (2002) also
finds that 160 particles are sufficient for percent-level shape
measurement.  To keep the corrections small while maintaining a 
moderate sample size, the Virgo data are cut at a mass of $10^{14} \hinv
\msol$, equivalent to 1467 particles for the \lcdm\  
case and 440 particles for the \tcdm\ case.  The numbers of halos in
each sample are listed in Table~\ref{tab:z0samples}.


\begin{deluxetable}{lcr}
\tablewidth{0pt} \tablecaption{Halo Samples at $z \se 0$ \tablenotemark{a}
\label{tab:z0samples}}
\tablehead{ \colhead{Model} & \colhead{$M_{min} [\hinv
      \msol]$} &  \colhead{N$_{cl}$} }
\startdata 
\lcdm$-$HV & $3 \times 10^{14}$ & 82967 \\
\lcdm$-$Virgo & $10^{14}$ & 353 \\
\tcdm$-$HV & $3 \times 10^{14}$ & 87121 \\
\tcdm$-$Virgo & $10^{14}$  & 703 \\
\enddata
\vspace* {-0.3truecm} 
\tablenotetext{a}{Mass-limited samples, $M_{200} > M_{min}$.}
\end{deluxetable}

\begin{figure} 
  \centering
  \vspace{-1.0truecm}  
  \epsscale{1.3}      
   \plotone{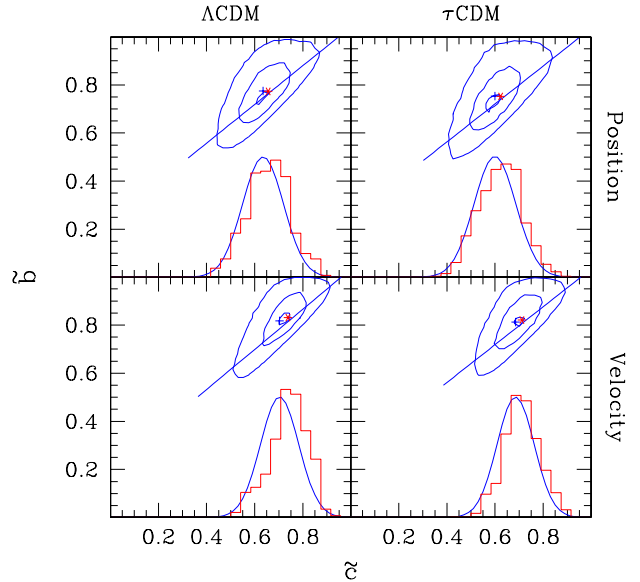}
  \vspace{-1.0truecm}  
   \caption{Contours of the joint probability distribution of 
axial ratios $p(\ctilde, \btilde)$ are shown with the minor-axis frequency
distribution $p(\ctilde)$ in the same format as 
Figure~\ref{fig:axial}. Left and right columns show
the \lcdm\  and \tcdm\ models while upper and lower rows give position and
velocity space distributions.  Contour levels in the joint pdf 
are drawn at the 10th, 50th, and 95th percentiles of the enclosed 
distribution.  Gaussian fits to the frequency distribution
of $\ctilde$ are shown for the HV 
simulations while histograms show $\ctilde$ values derived from the Virgo
simulations.  Crosses and asterisks mark the locations $(\langle \ctilde
\rangle, \langle \btilde \rangle)$ of the mean shapes for the HV and
Virgo models, respectively.
}                                              
   \label{fig:contour}                                                         
\end{figure}

The distribution of axial ratios at $z \se 0$ for 82,967 (\lcdm)
and 87,121 (\tcdm) halos are given in Figure \ref{fig:contour}.  The top row 
shows the distributions in position space while the bottom row
shows velocity space.  The location of the mean axial ratios  
($\langle \ctilde \rangle$, $\langle \btilde \rangle$) is depicted
by a cross for HV and an asterisk for Virgo data.  


Consistent with previous studies, we find galaxy cluster halos 
to be mostly prolate in shape, and somewhat rounder in velocity compared to
position space.  Table \ref{tab:peak} summarizes the $z \se 0$ shape
data.   The \lcdm\  HV simulation has modal values
(\caeq,\baeq)$_{peak}^{P}=(0.64, 0.76)$ and 
(\caeq,\baeq)$_{peak}^{V}=(0.72, 0.82)$ in position and velocity,
while the \tcdm\ model halos are more strongly ellipsoidal, with values
(\caeq,\baeq)$_{peak}^{P}=(0.59, 0.72)$ and 
(\caeq,\baeq)$_{peak}^{V}=(0.70, 0.80)$.

The frequency distribution of minor axis ratio $\ctilde$, computed by
integrating the joint pdf along the $\btilde$ axis, is well fit by a
Gaussian for the HV samples.  Distributions of
$\ctilde$ derived from the Virgo simulations, shown as histograms in
Figure~\ref{fig:contour}, are generally in good agreement with the HV
data.   The distribution of \lcdm\ minor axis ratios for position is
centered on $0.635$ and has measured dispersion $0.086$.  
When corrected for Poisson error, the estimated mean is slightly larger
$\langle \ctilde \rangle \se 0.647$ and the intrinsic
dispersion is estimated to be $0.076$.  
While the statistical uncertainty in the mean is in the
third significant digit, the level of systematic uncertainty is
certainly larger.  


\begin{deluxetable}{lccccc}
\tablewidth{0pt} \tablecaption{Mean Halo Shapes at $z \se 0$
  \tablenotemark{a} \label{tab:peak}}
\tablehead{ \colhead{Model-Space} 
& \colhead{$(\tilde{c}, \tilde{b})_{peak}$}
& \colhead{$ \langle \tilde{c} \rangle $} & \colhead{$\sigma_{\tilde{c}}$} &
\colhead{$\langle \tilde{b} \rangle $} & \colhead{$\sigma_{\tilde{b}}$} }
\startdata 
\lcdm$-$P & (0.635, 0.760) & 0.635 & 0.086 & 0.776 & 0.096\\ 
\lcdm$-$V & (0.719, 0.823) & 0.704 & 0.080 & 0.818 & 0.088\\
\tcdm$-$P & (0.593, 0.719) & 0.600 & 0.087 & 0.754 & 0.102\\
\tcdm$-$V & (0.698, 0.802) & 0.686 & 0.077 & 0.814 & 0.086\\
\enddata
\vspace* {-0.3truecm} 
\tablenotetext{a}{Mass-limited samples of Table~\ref{tab:z0samples}.}
\end{deluxetable}

A crude estimate of systematic error is given by the $0.012$ Poisson
bias correction for the mass-limited sample.  However, at the level of $0.01$,
a number of other effects on axial ratio come into play.  Foremost  
among them is the detailed definition of a halo: its location, scale,
and geometry.  We use the common spherical overdensity (SO) definition
of halos as spherical regions centered on local density peaks.  Other
viable approaches, such as friends-of-friends grouping or use of an
ellipsoidal boundary can systematically shift shape measurements by
many percent (Warren \etal 1992, 
Jing \& Suto 2002).  We leave it to future work to address such 
systematic effects in detail.  

The finding that halos are rounder in velocity space 
may be at least partly due the effects of ongoing mergers.  Mergers
will scatter in velocity space first, followed by mixing and 
relaxation of particle positions.  Another factor pushing in the same
direction is that the gravitational potential that drives the velocity
field is rounder than the density distribution.  

We do not attempt to formally fit the joint probability distribution
$p(\ctilde,\btilde)$, as doing so would requite a full deconvolution
of the effects of shot noise.  However, to give an indication of the
joint pdf shape, 
we locate the peak in the intermediate axis conditional probability
\begin{equation} 
p(\btilde | \ctilde) \ =
\ \frac{p(\ctilde,\btilde)}{p(\ctilde)}
\label{eq:jpf} 
\end{equation}  
and record the modal intermediate axis ratio $\btilde_{mod}$ as a
function of minor axis ratio $\ctilde$.  
The lines in each panel of Figure \ref{fig:contour} show fits
$\btilde_{mod}(\ctilde) \se r \ctilde + s$, with fit parameters given
in Table \ref{tab:peakline}.  


\begin{deluxetable}{lccc}
\tablewidth{0pt} 
\tablecaption{Intermediate Axis Modal Ridge Line
  $\btilde_{mod}(\ctilde)$ \tablenotemark{a} \label{tab:peakline} } 
\tablehead{ \colhead{Model} & \colhead{Component} & \colhead{$r$} 
& \colhead{$s$}}
\startdata 
\lcdm & Position & $0.81 \pm 0.02$ & $ 0.24 \pm 0.01$ \\ 
{  }  & Velocity & $0.84 \pm 0.03$ & $ 0.20 \pm 0.02$ \\
\tcdm & Position & $0.79 \pm 0.02$ & $ 0.25 \pm 0.01$ \\ 
{  }  & Velocity & $0.79 \pm 0.03$ & $ 0.25 \pm 0.02$ \\
\enddata
\vspace* {-0.3truecm}
\tablenotetext{a}{Fit to  $\btilde_{mod}(\ctilde) = r\ctilde +s$.}
\end{deluxetable}

\subsection{Position-Velocity Major Axis Alignment}

\begin{figure} 
  \centering     
   \vspace{-1.7truecm}
   \epsscale{1.3} 
    \plotone{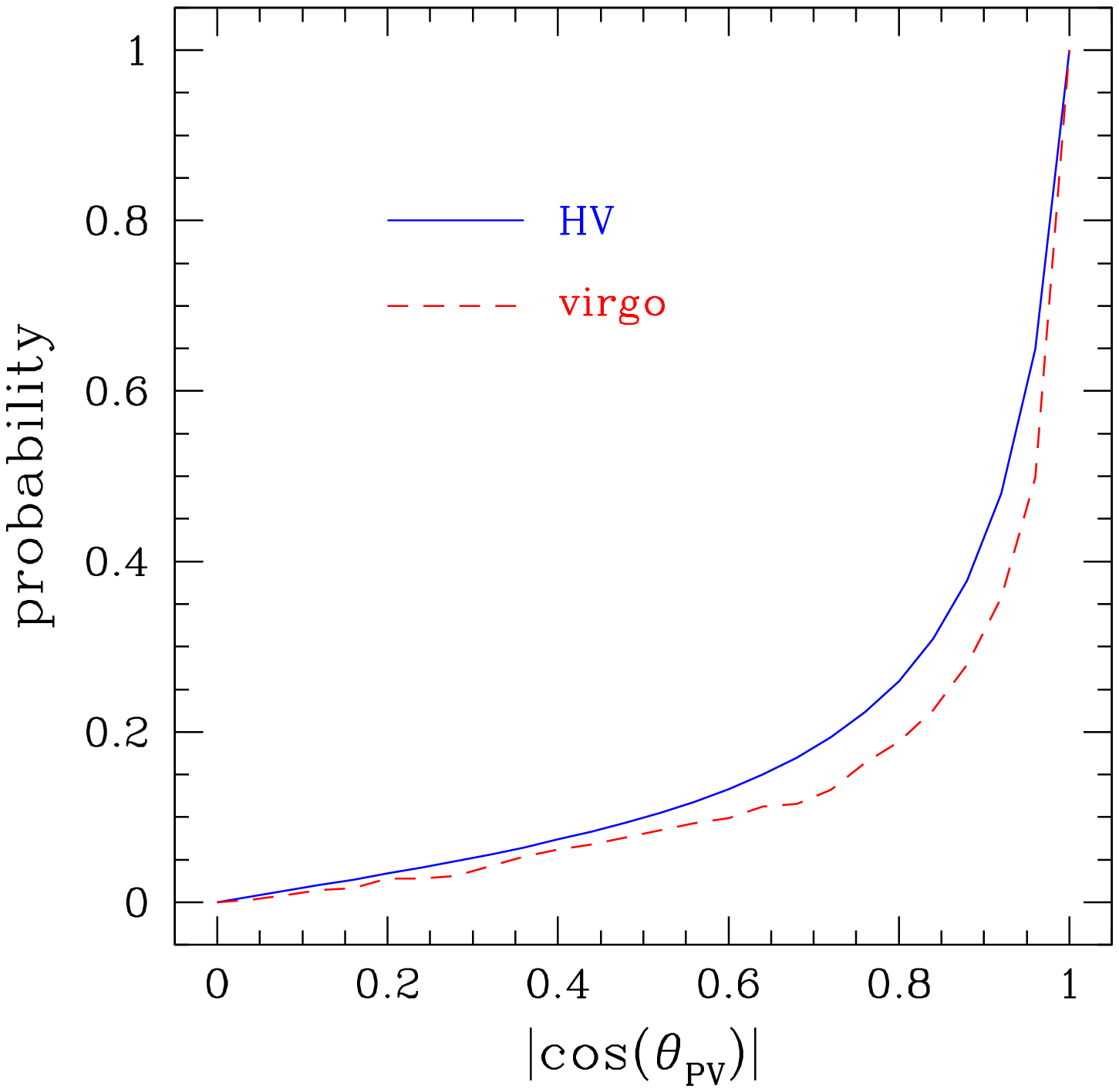}
   \vspace{-1.0truecm}
   \caption{Cumulative distributions of the alignment angle 
between the position and velocity major axes are shown for the
\lcdm\  HV clusters with $\Mtwoh \ge \Mcut$ 
(solid line) and \lcdm\  Virgo clusters with $\Mtwoh \ge 10^{14}
\hinv \msol$ (dashed line).  Half of the clusters in the \lcdm\ 
HV simulations have an alignment angle smaller than 22
degrees.}                                           
   \label{fig:thetaPV}                                                         
\end{figure}

Since elongated orbits drive both position and velocity
anisotropy, a correlation between both position and velocity
major axes is both expected and measured in simulations (Tormen 1997).
We quantify the alignment 
between position and velocity in a halo through the scalar product of
its major axis eigenvectors 
\begin{equation}
\cos(\theta_{PV}) \ = \ | \hat{a}^{P} \cdot \hat{a}^{V} |. 
\label{eq:costheta} \end{equation}
Figure \ref{fig:thetaPV} shows cumulative probability functions of 
this statistic for the \lcdm\ HV and Virgo models.  


We find a strong alignment signal: half of all halos have position
and velocity major axes aligned to better than 22 degrees for the HV
\lcdm\ model (21 degrees for \tcdm).  The higher resolution Virgo runs
show even stronger position-velocity alignment, with median values of
15 degrees for \lcdm\ and 14 degrees for the \tcdm\ simulation.  When
the same analysis is carried out using only strongly ellipsoidal
clusters ($\ctilde <  0.635$), the results do not significantly
change.  

Our results are consistent with a higher resolution study of
Tormen (1997), who finds from simulations of 9 dark 
matter halos resolved by 20,000 particles that the
position-velocity alignment angle is approximately 30 degrees. 

\begin{figure}
   \centering
   \vspace{-1.3truecm} 
   \epsscale{1.3}
   \plotone{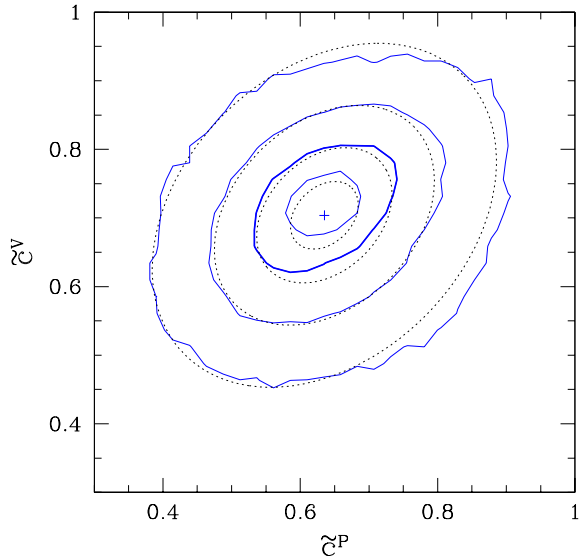} 
  \vspace{-0.7truecm} 
   \caption{Joint probability density of the position $\ctilde^P$ and
     velocity $\ctilde^V$ 
     minor axis ratios.  Light contours show the 16th, 50th, 84th and 99th
     percentiles of the enclosed distribution while dotted contours
     show the same for the ellipsoidal Gaussian fit described in the
     text.  The bold contour highlights the median of the joint distribution.  
}
   \label{fig:ca_PvsV}   
\end{figure}

To further define the relationship between position and
velocity space, we show the joint distribution of minor axis ratios
$(\ctilde^P,\ctilde^V)$ in Figure \ref{fig:ca_PvsV}.  The likelihood
is well fit by a Gaussian in an ellipsoidal coordinate $r$, where $r^2 =
(\frac{c_+}{\sigma_+})^2 + (\frac{c_{-}}{\sigma_{-}})^2$ and 
$c_{\pm} = \frac{1}{\sqrt{2}} [ (\ctilde^P - \langle \ctilde^P  \rangle) \pm
(\ctilde^V  - \langle \ctilde^V  \rangle) ]$ are principal
component directions centered on the one-component means of the
distribution.  This 
fit is shown by the dotted lines in Figure~\ref{fig:ca_PvsV}.  The
measured dispersions for \lcdm\ are  $\sigma_{+} \se 0.096$ and
$\sigma_{-} \se 0.069$. The bold contour shows the median of the
enclosed distribution.  


Since essentially all cluster detection methods use some power of
projected mass density as a defining signal, the spatial-velocity
alignments examined 
here will introduce an orientation bias in estimates of velocity
dispersions for clusters lying close to the sample detection
threshold.  The magnitude of this effect will depend on the specific
sample in question, but the general trend will be for the
line-of-sight dispersion to overestimate its isotropic counterpart, by
a fractional amount that could be $\sims 10\%$ for small samples of
moderate signal-to-noise detections.  Estimating the effect for 
particular observational surveys is best done using Monte Carlo
sample realizations, such as those performed for the 2MASS (Kochanek
\etal 2003) and SDSS (Miller \etal 2004; Wechsler \etal 2004) samples.

\subsection{Mass and Redshift Dependence}

Because mergers are directed along large-scale filaments in
the cosmic web, the birth of dark matter halos is an inherently
asymmetric process.  As a merger evolves, dynamical
relaxation will tend to drive a halo closer to isotropy.  One then 
expects that dynamically younger clusters will be more strongly
ellipsoidal than older ones.  Equating dynamical age with elapsed time
from a halo's formation epoch (defined, for example, using the
mass accretion history by Wechsler \etal 2002), we expect high mass
halos at a given epoch will be more elongated than those of lower
mass.  Similarly, at fixed mass, high-z halos should be more
ellipsoidal than their low-z counterparts of the same mass. 
  
\begin{figure*}
\begin{minipage}{180mm}
\epsfxsize=19.0cm \epsfysize=9.5cm
\centering
\vspace{-1.5truecm} 
\includegraphics[width=.48\textwidth]{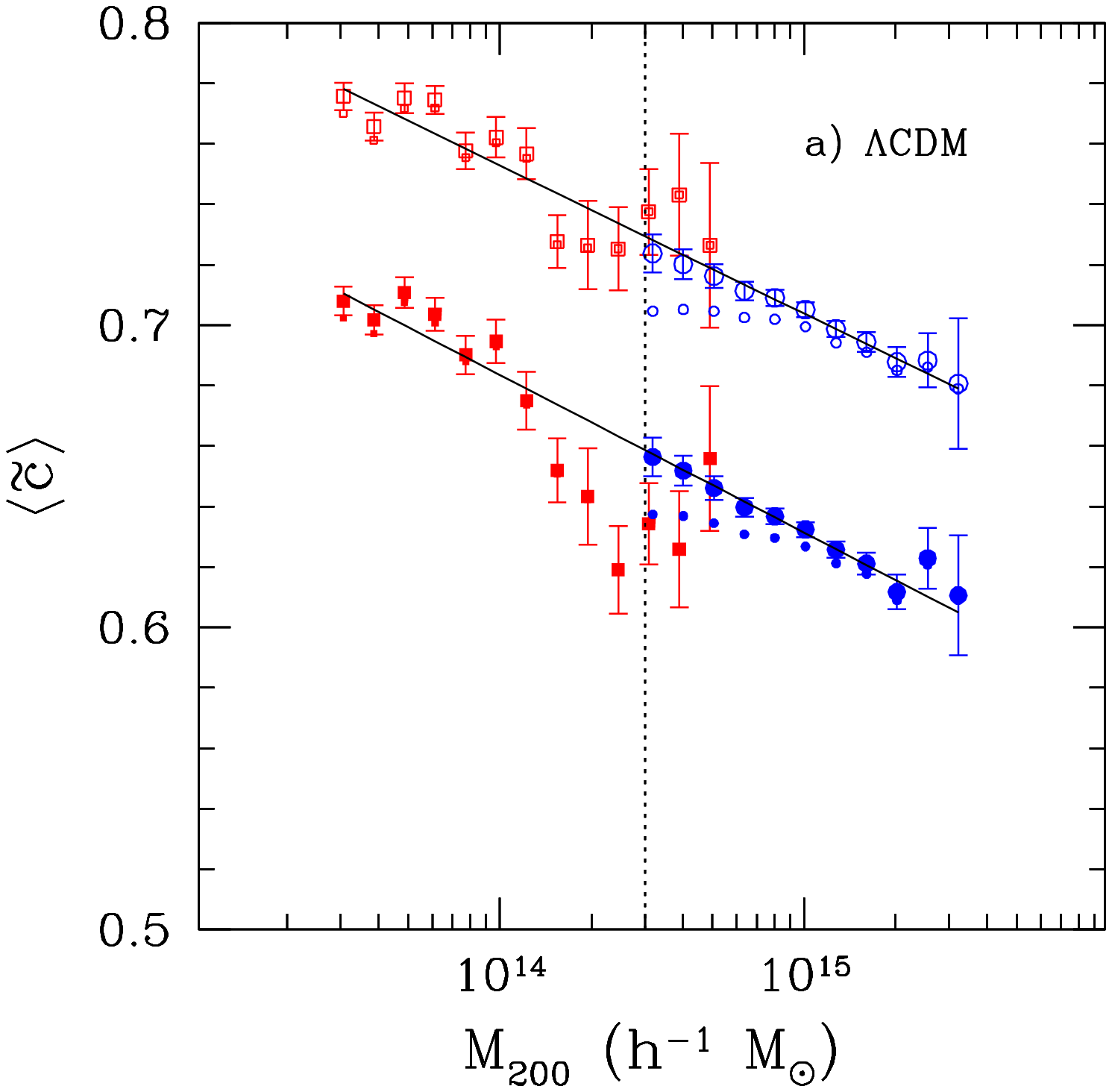}
\includegraphics[width=.48\textwidth]{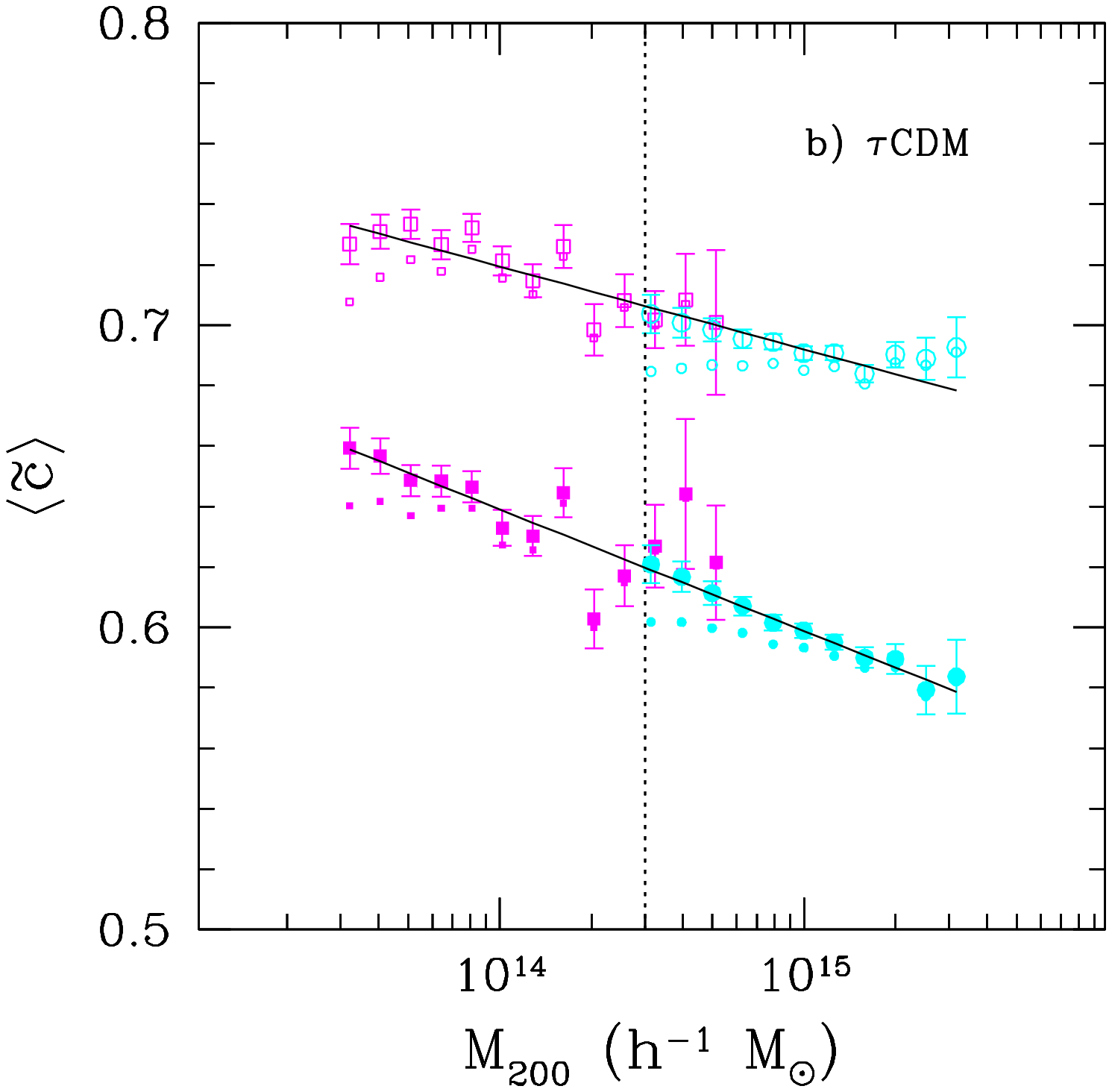}
\caption{The dependence of mean minor axis ratio on mass is
shown for halo samples at z=0 of the a) \lcdm\  and b)
\tcdm\ models.  The upper curve with open symbols give velocity shapes
while the lower curve with filled symbols show position.  Small
symbols show measures uncorrected for 
small-N effects while bigger symbols have Poisson corrections
applied, as discussed in the text.  Circles and squares are HV and
Virgo model results, respectively, and the vertical dotted line marks
the $\Mcut$ resolution limit of the HV models.  Solid lines show the
weak logarithmic 
dependence on mass, equation~(\ref{eq:mrdepend1}), with
parameter values given in Table~\ref{tab:ca_m}.  Higher mass clusters
are more strongly ellipsoidal than low mass clusters in both position
and velocity.}                
   \label{fig:ca_m}                                                      
\end{minipage}
\end{figure*}

With the large number of clusters in the $z\se 0$ sample,
we first investigate the mass dependence at the present epoch.  
Figure~\ref{fig:ca_m} shows the dependence of mean minor
axis ratio shapes in position (filled symbols) and velocity (open
symbols) on halo mass, using bins of width 0.1dex.  
Error bars on the points give the uncertainty in the mean.  Circles show
the HV data while squares show the Virgo data.  Small
symbols show the raw mean shape measurements while large symbols
correct for the bias in mean shape calibrated in section 3.1.  
There is generally good agreement
between the bias-corrected mean shapes of the HV models and the values 
measured directly from the higher resolution Virgo runs.  


\vskip10pt

The mass dependence favors more ellipsoidal halos at higher mass.  
The mean minor axis ratios of both the position and velocity shapes vary
by only $\sims 5$ percent as the mass changes by a factor of ten.  
For the \lcdm\ model, a fit to a logarithmic mass dependence using HV
data above $\Mcut$ and Virgo data below yields 
\begin{equation} 
\langle \ctilde^P \rangle  \ = \ (0.631 \pm 0.001) \ \bigl( 1-(0.023 \pm
  0.002) \ln M \bigr) ,
\end{equation} 
\begin{equation} 
\langle \ctilde^V \rangle \ = \ (0.704 \pm 0.001) \ \bigl( 1-(0.021 \pm
  0.001) \ln M \bigr) 
\label{eq:cofmLCDM} 
\end{equation} 
where $M$ is the halo mass measure $\Mtwoh$ expressed in units of
$10^{15} \hinv \msol$.  
Although these expressions are good fits to the shape 
dependence over two orders of magnitude in mass, the limit
$\ctilde \le 1$ requires that the shape at masses approaching $\Mtwoh
\lesssim 10^6 \hinv \msol$ must deviate this form.  Simulations of
smaller-scale structure will be needed to probe this regime. 

The \tcdm\ model halos are more ellipsoidal and display somewhat weaker
mass dependence than their \lcdm\ counterparts.  More vigorous
growth of linear perturbations in the \tcdm\ model drives a higher
frequency of halo mergers in this model (Lacey \& Cole 1994), and this
may explain why the \tcdm\ halos have a mean $\ctilde$ that is $\sims
0.03$ smaller than the \lcdm\ value.  The velocity shapes of the
\tcdm\ model present a more puzzling result, in that the logarithmic
slope $0.012 \pm 0.001$ is significantly shallower that the
\tcdm\ mass behavior.  We suspect that our correction for Poisson
noise may be inadequate for the velocities in this case 
(the mass resolution in the \tcdm\ Virgo run is poorer than
that of the \lcdm\ Virgo run, so corrections are larger for this
model).  The flat behavior of velocity shape for the few most massive
HV clusters also drives down the slope.  The slopes of position and
velocity shapes for the \lcdm\ model are consistent; massive halos in this
model have a fixed ratio of minor axis shapes
$\ctilde^P/\ctilde^V \se 0.896 \pm 0.003$.

The behavior at $z \se 0$ motivates the following form for the 
behavior of the mean minor axis ratio at arbitrary redshift 
\begin{equation} 
\langle \ctilde\rangle(M,z)  \ = \ \tilde{c}_{15}(z) \ (1-\alpha \ln M).
\label{eq:mrdepend1} 
\end{equation}  
Parameters at $z \se 0$ are listed in Table \ref{tab:ca_m}.  


\begin{deluxetable}{lccc}
\tablewidth{0pt} \tablecaption{Mass Dependence of Halo Minor Axis 
  Ratio\tablenotemark{a} \label{tab:ca_m}} 
\tablehead{ \colhead{Model} & \colhead{Component} &
\colhead{$\tilde{c}_{15}(0)$} & \colhead{$\alpha$} } 
\startdata 
\lcdm & Position & $0.631 \pm 0.001$ & $0.023 \pm 0.002$ \\ 
{  }  & Velocity & $0.704 \pm 0.001$ & $0.021 \pm 0.001$ \\ 
\tcdm & Position & $0.599 \pm 0.001$ & $0.017 \pm 0.001$ \\ 
{  }  & Velocity & $0.692 \pm 0.001$ & $0.012 \pm 0.001$ \\ 
\enddata
\vspace* {-0.3truecm} 
\tablenotetext{a}{From $z\se 0$ halo samples.}  
\end{deluxetable}

Turning to the redshift behavior, we first use the sky survey
data to verify that the mass slope $\alpha$ does not depend on
redshift.  Binning clusters in two broad redshift intervals, from $z
\se 0$ to $0.5$ and $0.5$ to 1, we fit the mass dependence of the mean
minor axis ratio and find consistency with the present-epoch slope.
For example, the \lcdm\ position minor axis ratio has 
$\alpha = 0.024 \pm 0.003$ and $0.025 \pm 0.003$ for the lower and
higher redshift ranges, respectively, both of which are consistent with
the $z \se 0$ value $\alpha = 0.023 \pm 0.001$.  

We characterize the redshift behavior of shape by fitting the
mass-intercept at $10^{15} \hinv \msol$ to a power law in expansion factor 
\begin{equation} 
\ctilde_{15}(z)  \ = \ \tilde{c}_{15,0} \ (1+z)^{-\varepsilon}.
\label{eq:c15z} 
\end{equation}  
In the sky survey samples, each halo of mass
$M$ and minor axis shape $\ctilde$ (bias corrected for Poisson noise)
at redshift $z$ contributes 
$\ctilde/(1-\alpha \ln M)$ to the estimate of $\ctilde_{15}(z)$.  We
use values of $\alpha$ measured at $z \se 0$.  

\begin{figure} 
  \centering    
  \vspace{-1.7truecm} 
   \epsscale{1.4}   
   \plotone{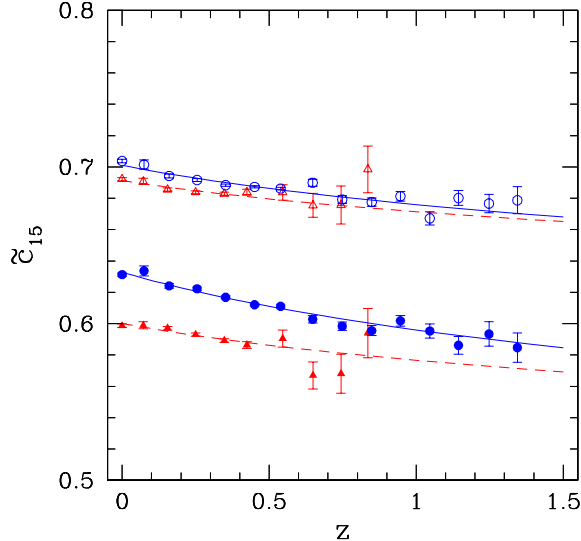}
  \vspace{-1.4truecm} 
   \caption{The redshift dependence of cluster shapes, expressed
in terms of the characteristic shape at $10^{15}\hinv \msol$, 
derived from the combined HV sky survey outputs of 
the \lcdm\  (circles) and \tcdm\ (triangles) models.  Filled and
open symbols are velocity and position values, respectively.
Lines are fits to equation~(\ref{eq:c15z}), with parameters given in
Table \ref{tab:ca_z}.}                          
   \label{fig:ca_z}                                                         
\end{figure}

Figure \ref{fig:ca_z} shows the results derived from the combined   
sky survey samples (PO, NO, MS, and VS) of halos with $\Mtwoh >\Mcut$,
binned in $\Delta z \se 0.1$ intervals.  A total
of 44,122 (\lcdm) and 19,813 (\tcdm) halos are above the mass limit.  
Filled symbols in Figure~\ref{fig:ca_z} are position while open show
velocity shapes.  Lines give fits to equation~(\ref{eq:c15z}) and the
fit parameters are listed in Table \ref{tab:ca_z}.  


The trend in redshift confirms
the expectation that high redshift halos are slightly more
ellipsoidal than their counterparts today.  The redshift dependence is
typically weak, $\varepsilon \ssim 0.05-0.09$, with the strongest trend
exhibited by the position minor axis of the \lcdm\ model.  

The present-epoch mean shapes at $10^{15} \hinv \msol$ are 
measured independently from the $z \se 0$ and sky survey
datasets.  The correspondence between $\tilde{c}_{15}(0)$ and
$\tilde{c}_{15,0}$ to nearly three significant digits implies that our
statistical uncertainties are well understood and that there are no
systematic differences at this level between the light-cone output
used to generate the sky surveys and the more common output produced
at fixed proper time.

\subsection{Comparisons to Previous Work}

Our shape results are generally consistent with previous work, but
different techniques for identifying halos, different shape
measurement conventions, and variations in assumed cosmology
complicate attempts at detailed comparison.   

The most recent work that is well aligned with our study is the N-body and
particle hydrodynamics simulations of \lcdm\ structure by  
Suwa \etal (2003).  Their cosmological model has slightly higher
normalization, $\sigma_8 \se 1$, but is otherwise identical to that
assumed in the HV and Virgo \lcdm\ runs.  They use a spherical halo
definition of somewhat larger radius (defined by an interior density
contrast of 200 with respect to the mean, not critical, mass density)
and find an average minor axis 
ratio $\langle \ctilde \rangle \se 0.62$ for clusters
more massive than $2 \times 10^{14} \hinv \msol$.  Their mass
resolution is an order of magnitude improvement over the HV
simulations, but their sample contains only 66 objects.  Floor \etal
(2003) use an annulus method to define projected shapes of halo from
several hydrodynamical and N-body simulations.  They find an average
projected axis ratio of 0.77 at $z \se 0$, a value consistent with the
mean intermediate axis ratio we find in three dimensions.


\begin{deluxetable}{lcccc}
\tablewidth{0pt} \tablecaption{Redshift Dependence of Halo Minor Axis
  Ratio\tablenotemark{a} \label{tab:ca_z}} 
\tablehead{ \colhead{Model} & \colhead{Component} &
\colhead{$N_{cl}$} & \colhead{$\ctilde_{15,0}$} &
\colhead{$\varepsilon$} } 
\startdata 
\lcdm & Position & 44122 & $0.633 \pm 0.001$ & $0.086 \pm 0.004$ \\ 
{  }  & Velocity & { }   & $0.701 \pm 0.001$ & $0.053 \pm 0.003$ \\
\tcdm & Position & 19813 & $0.600 \pm 0.001$ & $0.058 \pm 0.007$ \\
{  }  & Velocity & { }   & $0.691 \pm 0.001$ & $0.042 \pm 0.006$ \\
\enddata
\vspace* {-0.3truecm} 
\tablenotetext{a}{From combined sky survey samples.}  
\end{deluxetable}

Other published works employ an ellipsoidal region when defining halo 
shapes.  Warren \etal (1992) applied this technique to ground-breaking,
massively parallel simulations  and found a
distribution of minor axis ratios for galactic-scale halos that peaked 
at $\ctilde \ssimeq 0.6$, with dispersion $\sims 0.1$.  
Thomas \etal (1998) analyze four Virgo simulations, including the two
used here, and find mean minor axis ratio $0.50$ with dispersion
$\sims 0.15$ for a sample of 300 halos with mass limit $2 \times
10^{14} \hinv\msol$ in low-$\Omega_m$ cosmologies.  This shape result
is significantly more ellipsoidal than our mean value of $0.64$.  The
difference lies in the shape definition; we use moments of material
within a sphere while they use moments of material defined using a
percolation algorithm on a set of particles whose local densities lie above
a critical threshold.  The boundary constraint of our method 
will tend toward rounder 
measures while the latter method, because of the directional nature of the
percolation process and the pruning in local density, will allow more
strongly ellipsoidal values.  The 
degree of difference between the two methods can be large in extreme
cases.  For the same \lcdm\ Virgo simulations, the most extreme
position axis ratio measured by Thomas \etal (1998) is $<0.2$, while the
minimum value we derive is a factor two larger $0.4$.  Note that the
effect in the mean is much smaller, $\sims 20\%$.  

A more complete analysis involves measuring differential shapes as a
function of some scale parameter.  Warren \etal (1992) used an
iterative scheme that began within spheres of fixed physical radius
and found a high degree of correlation in direction and shape between
$10$ and $40 \hinv \mpc$ in a sample of galactic halos.  With much
higher resolution simulations, Jing \& Suto (2002) employ a sophisticated
approach that first measures a 
local density using a spherical kernel (as in SPH methods, Evrard
1988; Hernquist \& Katz 1989), then fits ellipsoids to particles
within some narrow range of local density.  

From $512^3$--particle simulations of a \lcdm\ cosmology, Jing \&
Suto (2002)
measure the distribution of minor 
axis shapes at a density $2500 \rho_c$ (corresponding to a radial
scale of roughly $\rtwoh/3$) in a sample of 2494 halos more massive than $6
\times 10^{12} \hinv \msol$.  They find mean $\langle \ctilde \rangle
=0.54$ and dispersion $0.11$ at $z \se 0$.  From analysis of twelve, 
high-resolution individual halo simulations, they find axial ratios
that are rounder at lower densities, with $\ctilde \ssim 
(\rho/\rho_c)^{-0.052}$.  Using this relation to roughly scale to the
radius used in this work, $\rho/\rho_c \se
80$ (equivalent to a mean interior density contrast of 200 for a $\rho
\spropto r^{-2.5}$ profile), results in $\langle \ctilde \rangle
=0.65$ for their mass-limited sample.
However, our measurement of the mass dependence of shape implies that a
second correction be made in
order to infer their expectations at our mass limit of $\Mcut$.  
Using equation~(\ref{eq:mrdepend1}), the 
result is an expected mean $\langle \ctilde \rangle =0.56$.  That the
HV sample mean of $0.64$ is significantly larger may again simply reflect the
different geometries being used in the two approaches, but this
hypothesis remains to tested.  Note that the scaled Jing \& Suto
(2002) result is $0.05$ larger than the mean derived by Thomas \etal (1998)
for halos more massive than $2 \times 10^{14} \hinv\msol$.  

Clearly, there is not a unique measure of absolute halo shape, and future
work is needed to more firmly establish the connections between
different approaches to shape measurement.  

The trends with mass and redshift of the mean minor axis ratio
presented in Figures \ref{fig:ca_m} and \ref{fig:ca_z} are in
qualitative agreement with Jing \& Suto (2002).  They find that halos
of higher mass (at fixed epoch) and higher redshift (at fixed mass)
tend to be more elongated.  Their fit 
to the mass dependence is equivalent to a value $\alpha \se 0.02$, in
good agreement with our finding of $0.023 \pm 0.002$.

Regarding observed trends of cluster shape with redshift, both
Plionis (2002) using 903 APM clusters and Melott et al. (2001) using
several optical and X-ray cluster samples find trends toward higher
ellipticity at increasing redshift.  However, Plionis et al. (2004) find a
trend of shape with cluster size that is opposite that seen in simulations.
From 1168 groups in the UZC-SSRS2 galaxy group catalog, Plionis et al. find
that poorer groups are more elongated than richer groups, with $85\%$ of poor
groups having $\btilde \lesssim 0.4$.   The discrepancy with the models may be
due to biases in the optical group catalogs or it may have a physical
origin, such as galaxies not fairly sampling the dark matter in low mass
systems. 


\section{SHAPE AND LARGE-SCALE STRUCTURE}

For the linear initial density field, a connection between cluster
shapes and large-scale structure was established by Bond, Kofman, \&
Pogosyan (1996), who showed that peaks separated by distances of order 
the mean inter-peak separation or smaller are likely to have strong
connecting filaments.  By directing mergers occurring on opposing ends,
filaments serve as a source of alignment for halo shapes.  Simulations
show that this alignment persists into the non-linear regime (van
Haarlem \& van de Weygaert 1993).  

In this section, we present mark correlations of shape, using two
measures employed previously by Faltenbacher \etal (2003).  We also
examine whether clusters that are members of superclusters 
have shapes that differ from the general population.  For the sake of
brevity, we present results for the \lcdm\ case only. 

\subsection{Spatial Correlations of Shapes}

Several observational studies have probed the
significance and length scale of spatial shape correlations of galaxy
clusters.  Plionis (1994), using galaxy positions in 637 Abell
clusters, presents evidence for nearest neighbor alignment 
extending to separations $15 \hinv \mpc$ at $\sims 2.5\sigma$
significance, with weaker alignment to $60 \hinv  
\mpc$. Fuller, West, \& Bridges (1999) used the brightest cluster 
galaxies (BCG) in poor MKW and AWM clusters and found significant
alignments for BCG-cluster pairs and BCG-nearest cluster pairs.  

Simulations demonstrate that alignments are to be expected.  Using
a sample of several hundred clusters more massive than $1.8 
\times 10^{14} \hinv \msol$ derived from a $512^3$--particle \lcdm\ 
simulation, Onuora \& Thomas (2000) detect an alignment signal for
pairs extending to $30 \hinv \mpc$, with the strongest signal for
nearest neighbors.  Like the position--velocity alignments of
Figure~\ref{fig:thetaPV}, they note that the signal is persistent and
changes little when only strongly elongated halos are used.

Faltenbacher \etal (2003; hereafter F03) examine 3000 clusters more 
massive than $1.4 \times 10^{14} \hinv \msol$ from a \lcdm\ 
simulation of a $500 \hinv \mpc$ region.  They introduce the use of
mark correlations to measure the behavior of alignment as a function
of radial scale, and find positive signal extending to $15 \hinv \mpc$
and $100 \hinv \mpc$ for two different alignment measures.   Hatton
and Ninin (2001) measure spatial alignments of halo pair angular momentums
that extend to similar scales. 

Following F03, we quantify orientation alignment with two different
measures.  Consider an ensemble of clusters pairs, each with members
$i$ and $j$ that have comoving spatial separation $r_{ij}$ between
$r$ and $r + dr$.  A basic alignment measure (termed ${\cal A}(r)$ by
F03) uses the scalar product of major-axis directions $\hat{a}$ 
\begin{equation} 
u(r) \ = \ \bigl\langle | \hat{a}_{i} \cdot \hat{a}_{j} | \bigr\rangle 
\label{eq:u_of_r} 
\end{equation}
where $\langle \ \rangle$ denotes an ensemble average over pairs of
separation $r$.  

A second, ``filamentary'' measure (${\cal F}(r)$ of
F03) compares halo major-axis orientation to pair separation direction 
\begin{equation} 
w(r) \ = \ \bigl\langle \frac{1}{2} \bigl ( |\hat{a}_{i} \cdot
  \hat{r}_{ij}| + | \hat{a}_{j} \cdot \hat{r}_{ij} | \bigr ) \bigr\rangle .
\label{eq:w_of_r} 
\end{equation} 
For random halo orientations, both measures have expectation value $u(r)
\se w(r) \se 0.5$.  We therefore use the deviations from this
expectation $\delta u(r) \se  u(r) - 0.5$ and $\delta w(r) \se  w(r) -
0.5$ as a measure of pairwise alignment.

\begin{figure} 
  \centering 
  \vspace{-1.8truecm}     
   \epsscale{1.45}         
  \plotone{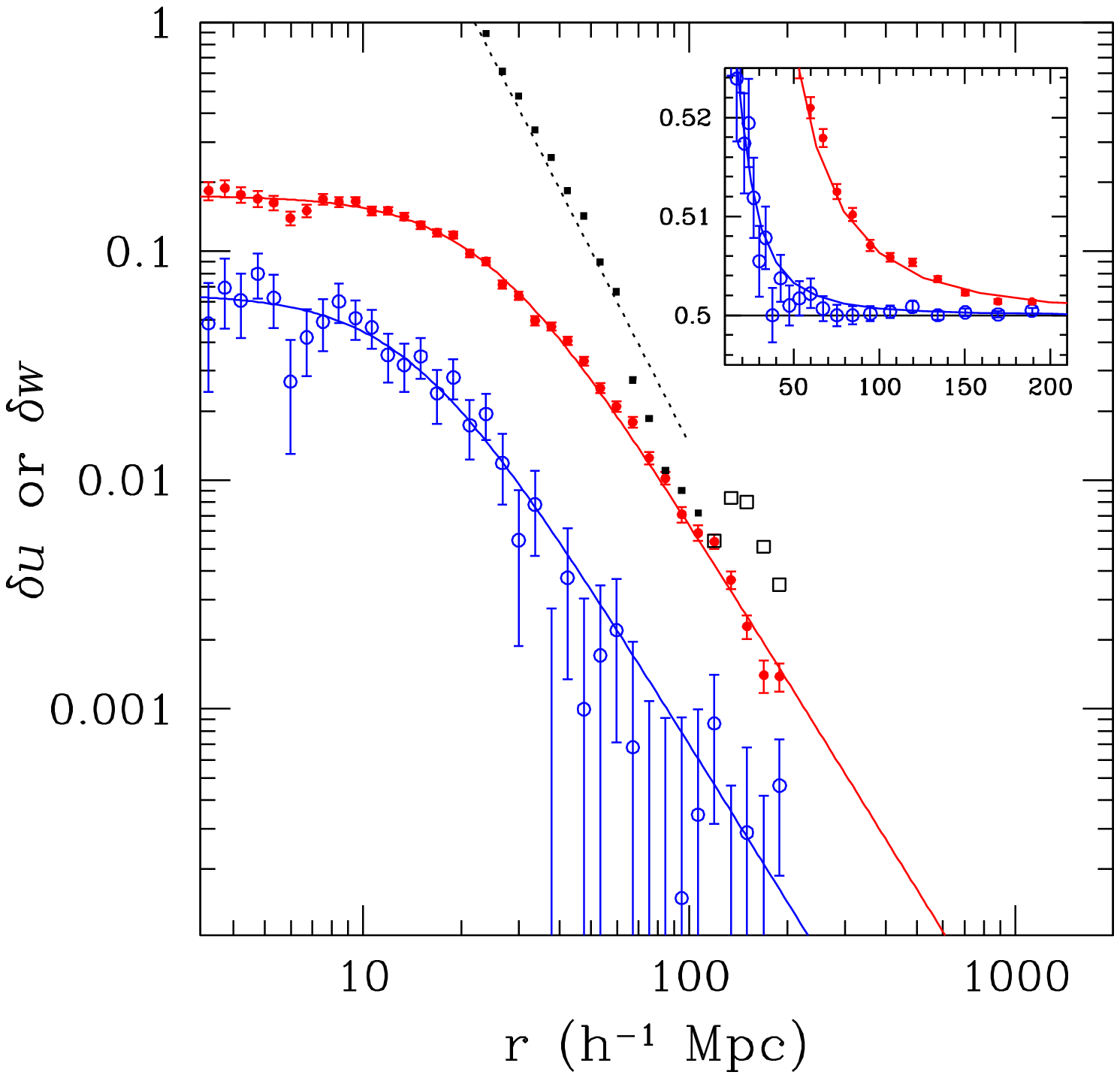}
  \vspace{-1.2truecm}     
   \caption{Mark correlation of cluster orientations as measured 
by the mean excess basic alignment $\delta u$ (open
circles) and the mean excess filamentary alignment $\delta w$
(filled circles) for the HV \lcdm\ sample with $\Mtwoh \ge \Mcut$ at
$z \se 0$.  Solid lines show the fits to equations~(\ref{eq:ufit}) and
(\ref{eq:wfit}).  Filled squares show the
sample's spatial correlation function in the regime where $\xi(r) > 0$
while open squares show $|\xi(r)|$ in  
the anti-correlated regime.  The dotted line is a rough fit of
$\xi(r)= (r/22 \hinv \mpc)^{-2.8}$ within the range $r \sim 20-60
\hinv \mpc$.}                 
   \label{fig:du_dw}                                                         
\end{figure}

Figure \ref{fig:du_dw} shows the \lcdm\ position-space alignment
signals at $z \se 0$ for the sample of 83,000 halos with masses
$\Mtwoh \ge \Mcut$.  Open circles show $\delta u(r)$ and filled 
circles give $\delta w(r)$.  Error bars are the uncertainty in the
binned mean values.  For comparison, squares show the 
two-point spatial correlation function $\xi(r)$ of the mass-limited
sample with correlation length $22 \hinv \mpc$ (Colberg \etal 2000).
Filled squares show positive spatial correlations while open squares
show $|\xi(r)|$ in the region of negative correlations $(\xi(r) < 0)$.
The inset shows $u(r)$ and $w(r)$ on a linear scale.  


The basic measure shows excess alignment out to scales $\sims 30 \hinv
\mpc$.  The filamentary statistic shows non-random halo
orientations extending to scales $200 \hinv\mpc$, nearly ten
times the correlation length and well into the
anti-correlated regime of the two-point function.  In both
cases, the excess alignment signal is well fit by a power law at large
radii that saturates at small $r$. The basic alignments follow 
\begin{equation} 
\delta u(r) \ = \ 0.065 \left( 1 + \left( \frac{r}{14 \hinv \mpc} 
\right) \right)^{-2.3} 
\label{eq:ufit}
\end{equation}
while the filamentary alignment is well fit by
\begin{equation} 
\delta w(r) = 0.175 \left( 1 + \left( \frac{r}{24 \hinv \mpc} \right)
\right) ^{-2.3}.
\label{eq:wfit} 
\end{equation} 

Our findings are generally consistent with those of F03, but our 
improved statistics and larger linear scale allow the first detailed
fits to the effect.  Although an analytic foundation for the specific
form of equations~(\ref{eq:ufit}) and (\ref{eq:wfit}) is not yet in
hand, we speculate 
that a solution may be found by applying the theoretical
machinery describing peaks in Gaussian random noise fields (Bardeen
\etal 1986; Bond \& Myers 1996).  

F03 note that the filamentary signal remains
strong in projection, but their analysis is optimistic in
that it assumes perfect knowledge of three-dimensional halo separations as
well as the projected three-dimensional halo shapes.  A more
appropriate treatment will require analysis of galaxy cluster samples
derived from mock catalogs (Kochanek \& White 2003; Wechsler \etal
2004). 

\subsection{Supercluster Members }

One might reasonably expect local cluster environment to play
a role in determining halo shapes.  In particular, since superclusters ---
groups of cluster-mass halos --- tend to have strong filamentary
morphology, one might anticipate that the formation dynamics of
supercluster members may lead to a distribution of shapes that is
biased toward more elongated systems.  

To address this question, we identify superclusters in the $z \se
0$ HV halo population 
by applying a percolation algorithm with linking length $23
\hinv\mpc$, one-third the $69 \hinv\mpc$ mean intercluster spacing of
the sample mass-limited at $\Mcut$.  We further require that each
supercluster have five or more halo members above the mass cutoff.
This selects the $8\%$ most strongly clustered halos.  


\begin{figure} 
  \centering     
  \vspace{-1.8truecm}
  \epsscale{1.4}  
  \plotone{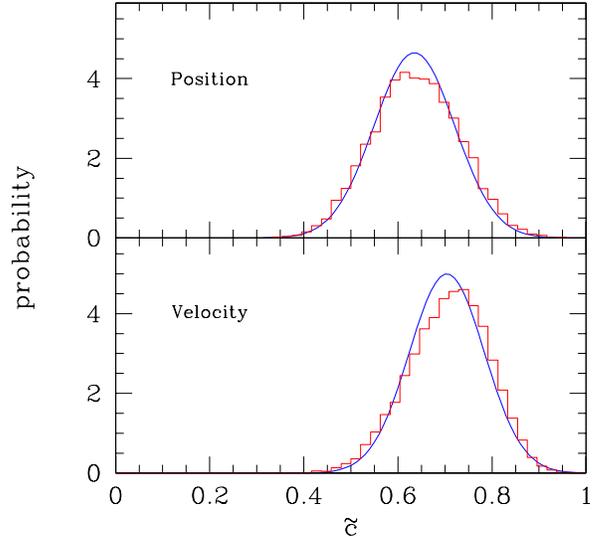}
  \vspace{-1.3truecm}
  \caption{Distributions of minor axis shapes for the 6683 halos of
the \lcdm\ supercluster population (histogram) compared with Gaussian fits
to the general population (solid lines) at $z \se 0$.  A minimum mass
of $\Mcut$ is employed.}
  \label{fig:supercl}        
\end{figure}  

Figure~\ref{fig:supercl} shows the distribution of minor axis ratios
in position and velocity for the supercluster members and the general
population.  
The axial ratios of the supercluster population have a nearly Gaussian
distribution with means and dispersions presented in
Table \ref{tab:supercl}.  Comparing supercluster 
members to the general population (Table \ref{tab:peak}), we find no
difference in 
mean values at the level of $0.01$ in axial ratio; both have $(\langle
\ctilde \rangle, \langle \btilde \rangle) =(0.64, 0.78)$ in position
and $(0.70,0.82)$ in velocity.  This finding lends support to the
picture in which the halo formation history is largely independent of
large-scale environment (Bower 1991; Sheth and Tormen 1999, 2004).  


\begin{deluxetable}{cccccc}
\tablewidth{0pt} \tablecaption{Supercluster Population Shapes\tablenotemark{a} \label{tab:supercl}}  
\tablehead{ \colhead{Model} & \colhead{N$_{cl}$} & \colhead{$\ctilde$} &
\colhead{$\sigma_{\ctilde}$} & \colhead{$\btilde$} &
\colhead{$\sigma_{\btilde}$} } 
\startdata 
Position & 6683 & 0.638 & 0.092 & 0.776 & 0.096\\ 
Velocity &  { } & 0.704 & 0.080 & 0.818 & 0.088\\
\enddata
\vspace* {-0.3truecm} 
\tablenotetext{a}{From \lcdm\ $z\se 0$ halo sample.}  
\end{deluxetable}

\section{CONCLUSIONS}

We use dark matter halos samples from billion--particle N-body
simulations to measure statistical properties of halo shapes. 
The main \lcdm\ samples consist of $83,000$ halos at $z \se 0$
and $44,000$ halos from sky survey output with masses 
$\Mtwoh \ge \Mcut$.  For each halo, a moment analysis of density and
velocity structure within $\rtwoh$ is used 
to define the principal axes of an equivalent ellipsoid.  
Higher resolution simulations and random realizations of isothermal
spheres demonstrate that systematic errors due to discreteness are
less than $0.02$ in mean axial ratio for the main HV samples.   

Massive halos have a Gaussian distribution of axis
ratios, with intrinsic dispersion $\sims 0.08$ and a 
mean minor axis ratio 
$\langle \ctilde\rangle(M,z)  \se \tilde{c}_{15,0} \, (1-\alpha \ln M)
\,(1+z)^{\varepsilon}$ that tends weakly toward rounder systems at
lower mass and redshift ($\alpha \sims 0.02$ and $\varepsilon \sims
0.09$ for position, see Tables \ref{tab:ca_m} and \ref{tab:ca_z}).  

Halos are rounder in velocity than in position space, a finding that
likely reflects more efficient scattering in velocity space and  
the rounder nature of the gravitational potential compared to the
density field.  The principal axes in 
position and velocity are strongly correlated;  50\% of halos 
have alignment angles smaller than 22 degrees.  We also provide a
Gaussian fit to the joint probability density of minor axis ratios in
position and velocity.  

We investigate mark correlations of cluster shape using two statistics
introduced by Faltenbacher \etal (2003).  We
measure significant excess alignment of halos extending to $30\hinv
\mpc$ for the basic measure and to $200 \hinv\mpc$ for the filamentary
statistic.  The latter is well fit as a function of scale by $\delta
w(r) = 0.175 [1 + (r/24 \hinv \mpc)]^{-2.3}$.  The filamentary
alignment should be detectable in large galaxy cluster surveys such as
SDSS, but precise comparison between theory and observation will
require a careful study of the connections between clusters observed in
redshift/color space and the underlying halo population. 

We also find that the distribution of halo shapes in supercluster regions
is indistinguishable from that of halos in general, showing that
shape is largely independent of large-scale environment.

The statistics presented here can be used to extend approximate methods
for creating non-linear representations of the matter distribution
based on the halo model description (Scocciamarro and Sheth 2002) by
incorporating an ensemble of ellipsoidal halos with
correlated orientations and aligned position and velocity ellipsoids.

\bigskip
S.K. acknowledges support from a General Electric Faculty for the Future
Fellowship, the Michigan Space Grant Consortium, the Michigan Center
for Theoretical Physics, and the Michigan Physics REU Program funded
by NSF.  A.E. acknowledges support from NASA ATP and NSF ITR
grants.  We are grateful to Y. Suto, Y. Jing, and M. Plionis for
useful comments on an early draft of this paper.













\end{document}